\documentclass[12pt,dvips]{article}
\usepackage{cite}
\@addtoreset{equation}{section} \makeatother

\def\beq{\begin{eqnarray}}
\def\eeq{\end{eqnarray}}

\def\lsim{\mathrel{\rlap{\lower3pt\hbox{\hskip0pt$\sim$}}
    \raise1pt\hbox{$<$}}}         %less than or approx. symbol
\def\gsim{\mathrel{\rlap{\lower4pt\hbox{\hskip1pt$\sim$}}
    \raise1pt\hbox{$>$}}}         %greater than or approx. symbol

\begin{document}

%%%%%%%%%%%%%%%%%%%%%%%%%%%%%%%%%%%%%%%%%%%%%%%%%%%%%%%%%%%%%%%%%
\title{
\vspace{1cm}
\Large\textbf{Strong CP Problem with $10^{32}$  Standard Model  Copies}
\vspace*{.5cm}
\author{ 
{\large \textbf{Gia Dvali$^{a,b}$\footnote{email: georgi.dvali@cern.ch, gd23@nyu.edu} and Glennys R. Farrar$^{b}$\footnote{email: gf25@nyu.edu} }}\\
\small \emph{$^a$CERN Theory Division, CH-1211, Geneva 23, Switzerland} \\
 \small \emph{$^b$Center for Cosmology and Particle Physics,}\\ \small \emph{Department of Physics, New York University}\\
 \small \emph{4 Washington Place, New York, NY 10003}}}

\date{}

\maketitle \thispagestyle{empty} \vspace*{.5cm}

\begin{abstract}
We show that a recently proposed solution to the Hierarchy Problem simultaneously solves the Strong CP Problem, without requiring an axion or any further new physics.  Consistency of black hole physics implies a non-trivial relation between the number of particle species and particle masses, so that with $\sim \, 10^{32}$ copies of the standard model, the TeV scale is naturally explained.  At the same time, as shown here, this setup predicts a typical expected value of the strong-CP parameter in QCD of $\theta\, \sim \, 10^{-9}$.  This strongly motivates a more sensitive measurement of the neutron electric dipole moment.
\end{abstract}

\newpage
\renewcommand{\thepage}{\arabic{page}}
\setcounter{page}{1}

\section{Introduction}

 The Standard Model suffers from  two major naturalness problems. These are the Hierarchy Problem 
 and the Strong CP Problem\cite{theta}.  Usually, new physics that is introduced to address one of these problems makes the other one even more severe. For example, low energy supersymmetry,  introduced to stabilize the Higgs mass,  brings at the same time many new arbitrary CP-violating  phases, which give additional contributions to strong CP-breaking.  And the Peccei-Quinn idea\cite{pq}, which beautifully solves the strong CP problem, requires a new intermediate scale of spontaneous  PQ symmetry breaking, which has to be explained, to set the value of the axion\cite{axion} decay constant. 
   
In this work, we show that the recently-suggested solution of the Hierarchy Problem which postulates the existence of $N \sim 10^{32}$ Standard Model copies\cite{largen}, automatically solves the Strong CP Problem as well, without  introducing  axions or any additional physics in the observable sector.  The new solution of the Hierarchy Problem relies on a bound from a consistency condition on Black Hole physics, relating particle masses $M$ and the number $N$ of particle species.  For large $N$ it reads\cite{largen} 
\begin{equation}
\label{bound}
M^2 \, \lsim \,  N^{-1} \, M_P^2 \, ,
\end{equation}
where $M_P$ is the Planck mass. It was further shown\cite{qg} that the gravitational cut-off of such a theory is at the scale
\begin{equation}
\Lambda \, \equiv \, {M_P\over \sqrt{N}} \, .
\end{equation}
This can be seen by a number of 
arguments, perhaps the simplest one being the observation that if black holes 
of size $\Lambda^{-1}$ could be treated as classical objects (as they normally would be treated in General Relativity  coupled to small number of species), they would evaporate in time $\lsim \Lambda^{-1}$, pointing to the inconsistency of the assumption of their classicality.  Thus, standard perturbation theory breaks down and gravity must change regime at scale $\Lambda$ (in agreement with perturbative renormalization of the Planck mass \cite{giagiga,veneziano}).  Additional evidence for this conclusion comes from the fact that $\Lambda$ is the maximal temperature of the system.   For a detailed discussion, we refer the reader to\cite{qg}.   

  The bound (\ref{bound}) offers the following `cheap' solution to the hierarchy problem.  
  All one has to do is to postulate the existence of 
 $N \sim 10^{32}$  copies of the Standard Model, that talk to each other only through gravity. 
 In the most economical scenario, all the copies are related by a certain permutation 
 symmetry, so there are no new low energy parameters in the theory and all the copies are strictly identical.   
   
 Although a very low energy observer from  each Standard Model replica may be puzzled by the smallness of the weak scale versus the Planck mass, the hierarchy  is  guaranteed by black hole physics.   If the cutoff of the theory could be arbitrarily high, this would create an unprecedented situation in which the Higgs mass is stabilized  without any new physics at that scale.  However
 with the knowledge that $\Lambda$ is a cut-off, everything falls into place.  The stabilizing ultraviolet  physics  will show up at energy scale $\Lambda$, in the form of strong gravitational physics and micro black holes.    
    
   We wish to show now, that the above model also solves the strong $CP$ problem.   The key point, to be explained below, is that the sum of all the $\theta$-angles is constrained by the ratio of the cutoff $\Lambda$ to the QCD scale. We have the consistency relation  
  \begin{equation}
\label{Ntheta}
\sum \, \theta_j^2 \, \lsim  \, \left ({\Lambda^4 \over \Lambda_{QCD}^4} \right )\, \approx \, { 1\over N^2} \, {M_P^4 \over \, \Lambda_{QCD}^4}. 
\end{equation}       
 Here $\theta_j, ~~j\, =\, 1,2,...N$, is the QCD $\theta$-parameter for the $j$-th copy of the Standard Model. Notice that because of the exact permutation symmetry, the value of the QCD scale is common (to leading order),   but not the value of the $\theta_j$ parameters, because they are integration constants.  Had we relaxed the exact symmetry requirement between the SM copies, the 
 bound (\ref{Ntheta}) would change to    
  \begin{equation}
\label{Ntheta1}
\sum \, \theta_j^2 \, \Lambda_{QCD_{j}}^4 \lsim  \, \Lambda^4, 
\end{equation}    
where $\Lambda_{QCD}^{(j)}$ refers to the  QCD scale of  the $j$-th copy. 
Notice that these are general relations, irrespective of whether we wish to solve the Hierarchy Problem or not.  The solution of the Hierarchy Problem corresponds to the choice $N \, \sim  \, 10^{32}$, which gives 
$\Lambda \, \sim$ TeV. Then, from (\ref{Ntheta}) it follows that, on average, 
the typical magnitude of $\theta$ is
\footnote{The same bound is obtained from the second consistency relation 
\begin{equation}
\label{Nthetal}
\sum \, \theta_j \, \lsim  \, \left ({\Lambda^4 \over \Lambda_{QCD}^4} \right )\, \equiv \, { 1\over N^2} \, {M_P^4 \over \, \Lambda_{QCD}^4}, 
\end{equation}    
due to the fact that because of random signs, in the large $N$ limit
$
\sum \, \theta_j \, \simeq \, \sqrt{N} \langle \theta \rangle\, $,
where the sum is approximated by a random walk with a step $\langle \theta \rangle$.} 
  \begin{equation}
\label{theta}
\langle  \theta  \rangle \, \equiv {1 \over N} \sqrt{ \sum \, \theta_j^2} \lsim  \, 10^{-9}\, . 
\end{equation}  

\section{$\theta$-Vacua as  Flux Vacua} 
\label{GGdual}
 In order to derive the relations (\ref{Ntheta}) and (\ref{Nthetal}), it is useful to  reformulate the strong-CP problem in the equivalent language of the electric field of  the QCD  Chern-Simons three-form (for a detailed discussion of  this formalism see\cite{axigauge} and references therein). 
  It is well known that in QCD the value of  $\theta$ is determined by the  
  vacuum expectation value  of the dual field strength $\langle {\rm Tr} G\tilde G \rangle$,  so that the 
  former can only vanish if the latter does and vice versa.     
   More precisely, for small $\theta$, we have\footnote{$\theta$ refers to the total angle, with both the 
   bare value and the contribution from the phase of the quark mass determinant included.}   
    \begin{equation}
\label{thetaggdual}
 \theta  \, \simeq  \,  {\langle {\rm Tr} G\tilde G \rangle \over \Lambda_{QCD}^4}.
\end{equation}  
The right hand side is the leading term in the expansion of an inverse periodic function (call it $f({\rm Tr} G\tilde G)$), which determines the dependence of the Lagrangian  on ${\rm Tr} G\tilde G$. The precise form of this function is unimportant for any of our conclusions, but we rely on the exact properties that it has zeros for ${\rm Tr} G\tilde G  \, = \, 0$ (thus guaranteeing $\theta \, = \, 0$ at these points, in accordance with the Vafa-Witten proof of the energy dependence on $\theta$\cite{vw}) and that its inverse periodicity guarantees  the invariance of physics under $\theta \, \rightarrow \, \theta \, + \, 2\pi$.  These are the only features we shall need.  
  
Evidently, an explanation for $\theta$ being small is completely equivalent to a demonstration that Tr$G\tilde G$ is small. Note that  the latter gauge-invariant can be rewritten as a dual four-form field strength of a Chern-Simons three-form in the following way (for simplicity,  here we shall work in units of the
QCD scale) 
 \begin{eqnarray}
\label{apsi}
  {g^2\over 8\pi^2} {\rm Tr} G\tilde{G} \, \equiv \, F\, \equiv \,  F_{\alpha\beta\gamma\delta}
\epsilon^{\alpha\beta\gamma\delta}, 
 \end{eqnarray}
where
 \begin{eqnarray}
\label{gluonfourform}
 F_{\alpha\beta\gamma\delta} \,\equiv \,  \partial_{[\alpha}C_{\beta\gamma\delta]}. 
 \end{eqnarray}
$C_{\alpha\beta\gamma}$ is a Chern-Simons three-form, which can be written in terms of gluon fields as
\begin{equation}
\label{qcdform}
C_{\alpha\beta\gamma} \, = \,  {g^2\over 8\pi^2}\, {\rm Tr} \left (A_{[\alpha}A_{\beta}A_{\gamma]}\,  - {3 \over 2}
A_{[\alpha}\partial_{\beta}A_{\gamma]}\right ).
\end{equation}
Here $g$ is the QCD gauge coupling, $A_{\alpha} \, = \, A^a_{\alpha}T^a$ is the gluon gauge field matrix, and $T^a$ are the generators of the $SU(3)$ group. 

 Under a gauge transformation, $C_{\alpha\beta\gamma}$ shifts as 
 \begin{equation}
\label{gauge}
C_{\alpha\beta\gamma}  \rightarrow  C_{\alpha\beta\gamma} \, + \, d_{[\alpha}\Omega_{\beta\gamma]},
\end{equation}
  with
\begin{equation}
\label{omega}
\Omega_{\alpha\beta}\, = \, A^a_{[\alpha}\partial_{\beta]}\omega^a,
\end{equation} 
where $\omega^a$ are the $SU(3)$ gauge transformation parameters. The four-form field strength
(\ref{gluonfourform})
is of course invariant under (\ref{gauge}) and (\ref{omega}). 

 From the discussion above, it is clear that  the $\theta$-parameter is just an expectation  value of the dual field strength $F$ in units of $\Lambda_{QCD}$, 
 and the $\theta$ vacua are nothing but vacua with different values of the constant three-form 
 electric flux.   Because in QCD there are no dynamical sources for $C_{\alpha\beta\gamma}$, its electric flux  cannot be screened or changed, either classically or quantum-mechanically.  This is why the $\theta$ vacua obey a superselection rule.  

Although it is not important for our discussion, we wish to briefly note that the interpretation 
in terms of the three-form gauge field is not just an useful  formality,  but has a clear physical meaning. 
It is known \cite{qcdform} that at low energies, the three-form $C_{\alpha\beta\gamma}$ becomes a massless gauge {\it field}, 
and supports  a long-range Coulomb-type constant electric flux.
The easiest way to see that $C_{\alpha\beta\gamma}$ does indeed behave as a massless field, 
is to note that  the correlator of the two  Chern-Simons
currents has a pole at zero momentum, which follows from the fact that 
the zero momentum limit of the  following correlator  
\begin{equation}
\label{corr}
 {\rm lim} _ {q \rightarrow 0} \, q^{\mu}q^{\nu} \, \int d^4x  {\rm e}^{iqx} \langle 0| T K_{\mu}(x)K_{\nu}(0)|0\rangle
\end{equation}
 (where  $K_{\mu} \, \equiv \, \epsilon_{\mu\alpha\beta\gamma} C^{\alpha\beta\gamma}$ is the Chern-Simons current) is non-zero, because the topological susceptibility of the vacuum is 
a non-zero number in pure gluodynamics. 
 Thus,  the three-form field develops a Coulomb propagator and can support a  
long-range electric flux.  In the absence of sources, this electric field is strictly static.   
The would-be sources for the three-form are two dimensional surfaces 
(axionic  domain walls or the 2-branes), but these are absent in minimal QCD. Since there are no sources,  there is no transition between vacua with different values of $F$.

 In other words,  at low energies the QCD Lagrangian contains a massless three-form field, and can be written as (suppressing combinatoric factors)
 \begin{equation}
\label{qcdtheta}
S  \, = \, \int \, d^4x \, ~~ \theta F \, + \, F^2 \, + \, ... \, .
\end{equation}
The above form has to be understood as an expansion of an inverse-periodic function $f(F)$ in powers of $F$,  an {\it exact}  property of which is that it has an extremum  at $F=0$; the explicit form is unimportant.    

This language gives a simple way for understanding the essence of the  strong CP problem, as well as 
for seeing  how $\theta$ determines the value of $F$.   By partial integration, the action can be written as (\cite{g98}; see also references therein) 
  \begin{equation}
\label{qcdtheta}
S  \, = \, \theta \int \, dX^{\alpha}\wedge dX^{\beta} \wedge dX^{\gamma} \, C_{\alpha\beta\gamma} \,  + \, \int d^4x \, F^2 \, + \, ...,
\end{equation}
where the first integral is taken over the $2+1$ dimensional world-volume of the space-time boundary surface, and $X^{\alpha}$ are its coordinates.   
 The equation of motion,
\begin{equation}
\label{fequation}
\partial_{\mu}\,  F^{\mu\nu\alpha\beta} \, = \, -\, \theta  \int \, dX^{\alpha}\wedge dX^{\beta} \wedge dX^{\gamma} \,  \delta^4(x\, - \, X),
\end{equation} 
then trivially implies that the solution that vanishes at the boundary is a step function, 
which away from the boundary  gives  $F \, = \, \theta$. 

 For example taking  the boundary to be a flat and a static surface located at $X^3 \, = \,  Z$,  the equation of motion becomes  
\begin{equation}
\label{feqstatic}
\partial_{\mu}\,  F^{\mu\nu\alpha\beta} \, = \, -\, \theta \delta(z\, - \, Z) \epsilon^{\nu\alpha\beta z},
\end{equation} 
which gives $F \, = \, \theta \, \Theta(Z \, - \,z)$, where $\Theta(z)$ is the step function. 
Notice that in the absence of the boundary, 
the solution is an arbitrary constant which would  be equivalent to effectively shifting $\theta$, but in both cases the physical parity-odd order parameter is the expectation value of $F$. 

The three-form 
language shows that there is a complete analogy between the $\theta$-vacua in  $3+1$-dimensional 
QCD, and the $\theta$-vacua of  $1+1$-dimensional massless electrodynamics\cite{1dim}. 
This is not surprising since a massless vector in $1+1$ and a massless three-form $3+1$ 
are both non-propagating, but both allow for a constant electric field.  
Interestingly, this analogy goes beyond the massless theory and also allows a unified description to be given of mass generation  
in the $1+1$ dimensional Schwinger model and  in $3+1$-dimensional  QCD with an $\eta'$ or axion,  purely in terms of topological entities\cite{jackiw}, although this is not relevant to our present application   
since we  are employing neither an axion nor a  massless quark.   
 
 \section{$\theta$-Vacua  in $10^{32}$ Copies of QCD} 
 
 Coming back to our agenda,  the connection
 (\ref{thetaggdual}) tells us that the physical parameter measuring the strong CP violation 
 is the Chern-Pontryagin electric flux  $F \, \equiv \, {\rm Tr} G\tilde G $.  Working with this quantity is convenient because we do not have to trace separately the different contributions in $\theta$, such as 
 its `bare' value and the contribution coming from the phase of the quark mass determinant, or possibly from some other sources. 
 At the end of the day, these  are all automatically summed up in the expectation value of $F$.   In QCD this fact is not of much help in understanding the smallness  of strong CP-breaking, since the natural value of $F$ is $\sim \, \Lambda_{QCD}^4$, implying $\theta \, \sim \, 1$.
  
 But, what if we have  $10^{32}$ copies of QCD?  The physical CP-odd order parameter  
 is now the sum over all the invariant fluxes.  Any universally-coupled physics, such as gravity, 
 will probe this collective flux, and not the individual entries.  The total electric  flux cannot exceed the cut-off of the theory.   In other words, just as the individual fluxes are bounded by the
 QCD scale,  the sum over all of them must be bounded by the cut-off of the theory.  This condition implies
 \begin{equation}
\label{Nflux}
\sum \, {F_j^2 \over \Lambda^4_{QCD}} \, \lsim  \, \Lambda^4 \, \equiv \,  {M_P^4 \over \, N^2},
\end{equation}    
and also 
 \begin{equation}
\label{Nfluxl}
\sum \, F_j \, \lsim  \, \Lambda^4 \, \equiv \,  {M_P^4 \over \, N^2} \, ,
\end{equation}    
giving  (\ref{Ntheta}) and (\ref{Nthetal}).  Similar bounds for higher power invariants give milder constraints. 
   
   Naively, it may seem that by requiring an exact symmetry between the Standard Model copies,  we can set all $\theta_j$ to be equal. This is not true, however, because the $\theta_j$'s are not fundamental parameters in the Lagrangian, but rather integration constants determined by the value of the Cern-Simons electric fluxes.  These fluxes cannot be restricted to be equal by symmetry, since they are solutions of the theory determining the different vacua, which satisfy a superselection rule.  Restricting the fluxes
  (equivalently the $\theta_j$'s) to be exactly equal, is equivalent to picking a very special  vacuum out of the continuum, but this does not make the other vacua to go away and the question 
 of why we do not find ourselves with $\theta_j$'s that are not exactly equal 
would still remain.  Therefore,  although there is a statistical upper bound 
on the $\theta_j$, we cannot claim that they should be equal.  
 
 Finally, we note that gravitational communication with the other copies cannot induce a substantial shift of $\theta$ in our sector. A potential source for such a shift is the kinetic mixing of Chern-Simons three-forms between the different copies: $F_jF_k$.  
Such a mixing is not forbidden by gauge symmetries, and can appear as a result of virtual black hole exchange. However, as shown in \cite{qg}, such operators can only appear for $j\neq k$ if they are suppressed by the scale $M_P^2\Lambda^2$, as opposed to the naive
$\Lambda^4$. The fundamental reason for such a strong suppression is that micro black holes cannot be universally coupled, so that inter-species transitions must be gravitationally suppressed\cite{qg}. It follows that the shift of $\theta$ in any given sector due to coupling with other sectors is
\begin{equation}
\label{shift}
\Delta \theta\, \sim \, \sum_j\, \theta_j {\Lambda^4_{QCD} \over M_P^2\Lambda^2} \, \sim\, 10^{-21}\,,
\end{equation}
and is un-observably small.
  
  \section{Conclusion} 
 We have shown that in the presence of $\sim 10^{32}$ copies of the Standard Model, 
 the maximum allowed value of the $\theta$ parameter in each given copy  {\it weakens} by a factor of $1/\sqrt{N}$.  Although the cumulative  CP-odd flux can be as large as the cut-off, the observable strong CP parameter in each  replica is small, and the strong CP problem is solved automatically. 
A very interesting feature of this mechanism is that the predicted value of $\theta$ is in the realm of the current observational upper bound.  If a neutron electric dipole moment is observed in next-generation precision experiments, it could be strong indirect evidence that whatever mechanism solves the Hierarchy Problem, is also responsible for the small size of strong CP violation.  

 Of course, the specific framework of $\sim 10^{32}$ copies of the Standard Model leaves many open questions, in particular the cosmological implications of so many species; some of these were studied in\cite{cosmo}.  However, whether or not this particular model is viable, simultaneously solving the Hierarchy Problem and the Strong CP Problem may be a general consequence of having a very large number of degrees of freedom.  Theories with very large numbers of degrees of freedom generically imply a low cut-off for gravity, suggesting that strong gravity effects such as micro black hole production will be observed in particle collisions above the TeV scale; this would provide another test of such scenarios.
 
  {\bf Acknowledgments}

We thank  Gregory Gabadadze, Martin~L\"uscher and Michele~Redi  for discussions. 
The work  is supported in part  by David and Lucile  Packard Foundation Fellowship for  Science and Engineering, and by NSF grants PHY-0245068 and PHY-0701451.

%%%%%%%%%%%%%%%%%%%%%% Bibliography %%%%%%%%%%%%%%%%%%%%%%%%%%%%%%%%%%%


\begin{thebibliography}{99}
\bibitem{theta}
C.G.~Callan, R.F.~Dashen and D.J.~Gross,  Phys. Lett. B63 (1976) 334;

R.~Jackiw and C.~Rebbi,  Phys. Rev. Lett. 37 (1976) 172. 

\bibitem{pq} 
 R.D.~Peccei and H.R.~Quinn, Phys. Rev. Lett. 38 (1977) 1440;
Phys. Rev. D16  (1977) 1791.

\bibitem{axion} 
S.~ Weinberg, Phys. Rev. Lett. 40 (1978) 223;

F.~Wilczek, Phys. Rev. Lett. 40 (1978) 279

\bibitem{largen} 
 G.~Dvali, arXiv:0706.2050 [hep-th] 
 
 \bibitem{qg} 
 G.~Dvali and M. Redi, arXiv: 0710.4344 [hep-th]
 
 \bibitem{giagiga}  G.~Dvali and G.~Gabadadze, Phys.Rev.D63:065007,2001.

\bibitem{veneziano} G.~Veneziano,  JHEP 0206:051,2002. 

\bibitem{axigauge} G.~Dvali, hep-th/0507215.

\bibitem{vw}
C.~Vafa and E.~Witten,  Phys. Rev. Lett. 53 (1984) 535.

\bibitem{qcdform} M.~L\"uscher, Phys. Lett. B78 (1978) 465.

\bibitem{g98}  G.~ Gabadadze,  Phys.Rev.D58:094015,1998; Nuc. Phys. B {\bf 552}, 194 (1999).

\bibitem{1dim} S.~Coleman, R.~Jackiw, L.~Susskind,  Ann. Phys. 93 (1975) 257.

\bibitem{jackiw} 
G.~Dvali, R.~Jackiw and S.-Y.~ Pi, {\it Phys. Rev. Lett.} {\bf 96} (2006) 08162.

\bibitem{cosmo} Y. Watanabe, E. Komatsu,   arXiv:0711.3442 [hep-th]

\end{thebibliography}
\end{document}